\documentclass[aip,jap,reprint,showpacs,amsmath,amssymb,superscriptaddress]{revtex4-1}
\usepackage{graphicx}
\usepackage{dcolumn}
\usepackage{bm}
\usepackage{color}

\begin{document}

\title{Engineering the Quantum Anomalous Hall Effect in Graphene with Uniaxial Strains}
\author{G. S. Diniz}
\email{ginetom@gmail.com}
\affiliation{Institute of Physics, University of Bras\'ilia, 70919-970, Bras\'ilia-DF, Brazil}
\author{M. R. Guassi}
\address{Institute of Physics, University of Bras\'ilia, 70919-970, Bras\'ilia-DF, Brazil}
\author{F. Qu}
\affiliation{Institute of Physics, University of Bras\'ilia, 70919-970, Bras\'ilia-DF, Brazil}
\affiliation{Department of Physics, The University of Texas at Austin, Austin, Texas 78712, USA}

\date{\today}

\begin{abstract}
We theoretically investigate the manipulation of the quantum anomalous Hall effect (QAHE) in graphene by means of the uniaxial strain. The values of Chern number and Hall conductance demonstrate that the strained graphene in presence of Rashba spin-orbit coupling and exchange field, for vanishing intrinsic spin-orbit coupling, possesses non-trivial topological phase which is robust against the direction and modulus of the strain. Besides, we also find that the interplay between Rashba and intrinsic spin-orbit couplings results in a topological phase transition in the strained graphene. Remarkably, as the strain strength is increased beyond approximately 7\%, the critical parameters of the exchange field for triggering the quantum anomalous Hall phase transition show distinct behaviors - decrease (increase) for strains along zigzag (armchair) direction. Our findings open up a new platform for manipulation of the QAHE by an experimentally accessible strain deformation of the graphene structure, with promising application on novel quantum electronic devices with high energy efficiency performance.
\end{abstract}

\pacs{73.22.Pr,73.43.Cd,75.50.Pp,61.48.Gh,77.65.Ly}

\maketitle
\section{Introduction}

Graphene - a truly two-dimensional material, composed only by covalently bonded carbon atoms in a honeycomb lattice, has been attracting the attention of the scientific community since its first well succeeded realization \cite{Novoselov22102004}. Most of its interests are due to the unusual electronic, thermal and nanomechanical properties with potential applications in wide variety of fields, for instance: spintronics \cite{MRS:8753723}, Majorama fermions proposals for quantum computation \cite{1402-4896-2012-T146-014013}, electron optics \cite{PhysRevB.87.075420}, photonics \cite{10.1021/nn300989g} and many others.

The prospect of using graphene in spintronic devices relies on the understanding of spin-orbit coupling (SOC). Two different SOC contributions are present in graphene: (i) \emph{extrinsic} Rashba SOC, originated from interactions with the substrate, electric field or curvature \cite{Huertas,Sancho,PhysRevLett.100.107602} and (ii) the \emph{intrinsic} SOC (ISO) originated from the carbon intra-atomic SOC, which is proposed to give rise to an insulating bulk electronic state that supports the transport of charge and spin in conducting edge states along the sample boundaries. This emerging time reversal invariant electronic state is the so-called quantum spin-Hall phase (QSH) \cite{Haldane,Kane,PhysRevLett.95.226801,PhysRevLett.101.146802,RevModPhys.82.3045,nl401147u}.

Since SOC holds time reversal symmetry (TRS), the spin current in a given edge is robust against scattering induced by nonmagnetic impurities \cite{Kane,PhysRevLett.95.226801}. Nevertheless, it might be suppressed by a magnetic field due to breaking down of TRS. A suppression of QSH is compensated by an emerge of another state of matter - \emph{quantum Hall state} - characterized by a precisely quantized Hall conductance. One of the grand challenges of the quantum Hall effect (QHE) is controlling spin-dependent properties without using external magnetic fields. To do so, one can resort to the SOC combined with an internal magnetization, which is responsible for breaking TRS. This kind of study has given rise to new physical insights in established phenomena, such as the quantum \emph{anomalous} Hall effect, in which one of the spin channels in the QSH state is suppressed by the sample magnetization, that may be experimentally achieved through magnetic atom doping in the sample \cite{PhysRevB.85.115439,RevModPhys.82.1959,Chang12042013}.

It is known that strains can be introduced on graphene either intentionally by tensions at the sample-leads contacts on suspended graphene devices \cite{NanoBao}, or naturally induced by the substrate deformation in which the graphene is deposited on top \cite{apl3463460,PhysRevB.79.205433,nn800031m,nn800459e,apl4821364}. Since the strain stretches the hexagonal lattice of the graphene out of equilibrium, with high degree of tunability in a diverse type of applied mechanical strain \cite{Lee18072008,PhysRevB.78.075435}, it may directly alter one of the degrees of freedom: namely the pseudospin. Consequently, it induces changes of SOC as well as QAHE.  Therefore, strains can offer an ideal opportunity to create new paradigms of QAHE intentional control in graphene \cite{PhysRevB.88.045425}.

Motivated by the prospects of external manipulation of the QAHE in graphene by strains \cite{nl901448z,PhysRevB.88.045425,PhysRevB.78.205433,PhysRevLett.109.066802}, we present in this paper results of the microscopic study of the QAHE in graphene under uniaxial strains. For this purpose, we have theoretically explored the dependence of electronic structure, topological and transport properties upon the orientation and modulus of uniaxial strain, in the presence of Rashba, ISO and an exchange field interaction.

\section{Theoretical Model}
We consider a graphene honeycomb lattice in the $x$-$y$ plane in presence of uniaxial strains with homogeneous Rashba SOC, ISO and exchange field interaction. The graphene is described by a $\pi$-orbital orthogonal tight-binding model with nearest-neighbor hopping. The Hamiltonian for this system in the real space is described by
\begin{eqnarray}
\label{H1}
H = H_{KM} + H_{M},
\end{eqnarray}
where

\begin{eqnarray}
H_{KM} &=& \sum_{\langle i,j\rangle,\sigma}t_{ij}c_{i\sigma}^{\dagger}c_{j\sigma} + i\lambda_{R}\sum_{\langle i,j\rangle}\hat{z}\cdot (\vec{s}\times \vec{\delta}_{ij})c_{i}^{\dagger}c_{j}\\\nonumber
       &+&
\dfrac{2i}{\sqrt{3}}\lambda_{so}\sum_{\langle\langle i,j\rangle\rangle}c_{i}^{\dagger}\vec{s}\cdot \left(\vec{d}_{kj}\times \vec{d}_{ik}\right)c_{j}\\
H_{M}  &=& M\sum_{i;\sigma,\sigma^{\prime}}c_{i\sigma}^{\dagger}s_{\sigma\sigma^{\prime}}^{z}c_{i\sigma^{\prime}}.
\end{eqnarray}
Here $H_{KM}$ is the Kane-Mele model, in which the first term is kinetic energy, the second and third terms are the Rashba SOC and ISO \cite{Kane}, respectively. The operators $c_{i\sigma}^{\dagger}$/$c_{i\sigma}$ creates/anihilates an electron at site \emph{i} with spin $\sigma$ ($=\uparrow, \downarrow$) and hopping amplitude $t_{ij}=t_{i}=t_{0}e^{-3.37\left(\delta_{i}-1\right)}$, where unstrained graphene hopping $t_{0}\approx-2.9$eV \cite{Neto} and the deformed lattice distances $\vec{\delta}_{i}$ are related to the relaxed ones $\vec{\delta}_{i}^{0}$ by $\vec{\delta}_{i}=\left(1+\epsilon\right)\vec{\delta}_{i}^{0}$, with $\vec{\delta}_{1}^{0}=(0,-1)$, $\vec{\delta}_{2}^{0}=1/2(\sqrt{3},1)$ and $\vec{\delta}_{3}^{0}=1/2(-\sqrt{3},1)$. The vectors $\vec{d}_{ij}$ points from $j$ to $i$, which for the ISO with coupling parameter $\lambda_{so}$ connects the next nearest-neighbors through $k$, $\vec{s}$ are the Pauli spin matrices. The Rashba SOC parameter $\lambda_{R}$ is proportional to the electric field applied perpendicular to the $x$-$y$ plane of the graphene \cite{Zarea,parameters}. The term $H_{M}$ corresponds to the uniform exchange field with strength $M$ responsible for breaking TRS of the system \cite{PhysRevB.83.155447}. In the elastic regime, the uniaxial strain tensor $\epsilon$ that relates the relaxed lattice vector to the strained ones can be written as \cite{PhysRevB.80.045401}
\begin{eqnarray}
\label{strain}
\epsilon=
\varepsilon\left(\begin{array}{cc}
\cos^{2}\theta -\nu\sin^{2}\theta & (1+\nu)\cos\theta\sin\theta\\
(1+\nu)\cos\theta\sin\theta & \sin^{2}\theta -\nu\cos^{2}\theta\\
\end{array}\right),
\end{eqnarray}
where $\nu=0.165$ is the Poisson's ratio value known for graphite \cite{PhysRevB.80.045401}, $\theta$ denotes the angle along which the strain is applied with respect to the $x$-axis in the direct lattice coordinate system, with $\vec{a}_{1}^{0}=a/2(2\sqrt{3},0)$ and $\vec{a}_{2}^{0}=a/2(\sqrt{3},3)$ and $\varepsilon$ is the strain modulus \cite{PhysRevB.80.045401}. We set the unstrained C-C distance $a$ to be unity for simplicity.

By making Fourier transformation of Eq. \ref{H1} and taking into account the effect of strain presented in Eq. \ref{strain}, we obtain a $4\times4$ matrix $H(\vec{k})$ in the momentum space, that can be numerically diagonalized for each crystal momentum to obtain the energy eigenvalues and eigenvectors.

To identify the topological properties of the Dirac gap and study the origin of QAHE, we have calculated the Berry curvature of the n$th$ bands $\Omega_{xy}^{n}(k_{x},k_{y})$ using the Kubo formula
\begin{equation}
\Omega_{xy}^{n}(k_{x},k_{y})=-\sum_{n^{\prime}\neq n}\dfrac{2Im\langle \Psi_{nk}\vert v_{x}\vert \Psi_{n^{\prime}k}\rangle\langle \Psi_{n^{\prime}k}\vert v_{y}\vert \Psi_{nk}\rangle}{(\omega_{n^{\prime}}-\omega_{n})^{2}},
\label{Berry}
\end{equation}
where $\omega_{n}=E_{n}/\hbar$ with $E_{n}$ the energy eigenvalue of the n$th$ band and $v_{x(y)}={\hbar}^{-1}\partial H/\partial k_{x(y)}$ is the Fermi velocity operator. When the Fermi level lies within the bulk gap, i.e. in the insulating regime, according to the Kubo formula, the corresponding Hall conductance is quantized as $\sigma_{xy}=\mathcal{C}e^{2}/h$. Where $\mathcal{C}$ is defined as the Chern number and can be calculated by \cite{PhysRevLett.49.405}
\begin{equation}
\mathcal{C}=\dfrac{1}{2\pi}\sum_{n}\int_{BZ}d^{2}k \Omega_{xy}^{n},
\label{H4}
\end{equation}
where the summation is taken over all the occupied states below the Fermi level and the integration is carried out over the whole first Brillouin zone.

Since the Berry curvatures are highly peaked around the Dirac points $\textbf{K}$ and $\textbf{K}^{\prime}$ \cite{nl300610w}, then a low energy approximation can be used in the calculation of the Chern number \cite{PhysRevB.82.161414}. This allows us to derive an effective tight-binding Hamiltonian of the strained graphene, by expanding $H(\vec{k})$ at the vicinity of the strain-shifted Dirac points, i.e., $\vec{k}=\eta \textbf{K} +\vec{q}$, where $\eta=\pm1$ labels the two valleys and $\vec{q}=(q_x,q_y)$ is a small crystal momentum around $\eta \textbf{K}$ (see Appendix. \ref{alowene} for further details). The validity of the low energy approximation requires the strain modulus to be upper limited, such that does not go beyond the threshold of an appearance of a band gap, thus the band is still linear and gapless at the strain-shifted Dirac points, in the absence of SOCs and exchange field interactions \cite{PhysRevB.80.045401}. This condition is fulfilled by the relation on the strain-dependent hopping parameters $\vert t_{1}-t_{2}\vert \le t_{3} \le \vert t_{1}+t_{2}\vert$, where $t_{i=1,2,3}$ are the hopping along each C-C bond \cite{PhysRevB.74.033413}. Under this condition, we calculate the Chern number using the following equation,

\begin{equation}
\mathcal{C}=\dfrac{1}{2\pi}\sum_{K,K^{\prime}}\sum_{n=1,2}\int_{-\infty}^{{\infty}}dq_{x}dq_{y} \Omega^{n}_{xy}(q_{x},q_{y}).
\label{H4}
\end{equation}

It is interesting to mention that in the above integral, a momentum cutoff is set around each valley for which the Chern number calculation is guaranteed to converge.

\section{Numerical Results and Discussions}
\begin{figure}[!h]
\centering
\includegraphics[scale=0.35]{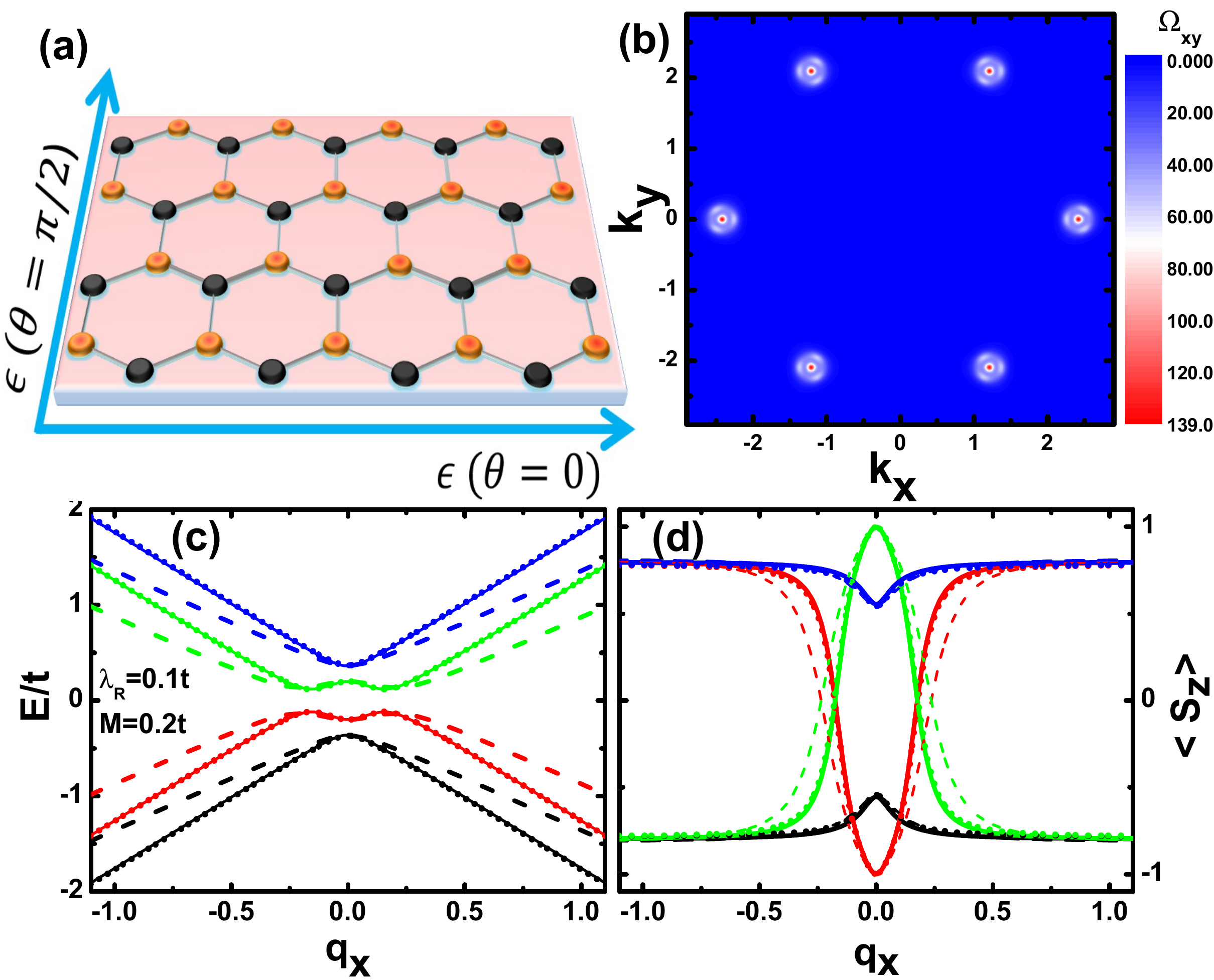}
\caption{(a) Schematic diagram showing the graphene above a substrate. The arrows indicate the strain direction: $\theta=0$ (zigzag direction) and $\theta=\pi/2$ (armchair direction). (b) Berry curvature of upper valence band for the unstrained graphene. (c) Energy dispersion of the graphene in presence of Rashba SOC and exchange coupling along $q_{x}$ for $q_{y}$=0. (d) Spin expectation $\langle S_{z}\rangle$ of an electron in correspondent states shown in (c). In the panels (c)-(d): solid, dashed and dotted lines are for $\varepsilon$=0 and $\varepsilon$=0.1 along $\theta=0$, and $\theta=\pi/2$, respectively. $\lambda_{R}$=0.1$t$, $\lambda_{so}$=0 and $M$=0.2$t$ have been used in panels (b)-(d).}
\label{grafico1}
\end{figure}

In Fig. \ref{grafico1} (a), we schematically show the sample which is used in our model calculation - the graphene subjected to an uniaxial strain whose directions are indicated by the arrows: $\theta$=0 (along zigzag direction) and $\theta$=$\pi/2$ (along armchair direction). Fig. \ref{grafico1} (b) shows the contour plot of the Berry curvature distribution of upper valence band, in the momentum space for the unstrained graphene. One can note that sharp peaks of the Berry curvature are localized close to the Brillouin zone corners and share the same sign. Fig. \ref{grafico1} (c) plots the influence of magnitude and orientation of the strain on the energy dispersion along $q_{x}$ for $q_{y}$=0, with $\lambda_{R}$=0.1$t$, $\lambda_{so}$=0 and exchange field $M$=0.2$t$. The spin expectation value $\langle S_{z}\rangle$ associated with the correspondent states illustrated in (c) is shown in (d). The $\langle S_{z}\rangle$ reaches its maximum value when the momentum approaches to $q_{x}$=0. This indicates a weak spin mixture for the lowest conduction and highest valence bands close to the Dirac points. In the case of $\theta=\pi/2$, and same strength $\varepsilon$=0.1 (dotted lines) the energy dispersion shows no effect in comparison with the unstrained case, in fact this behavior is due to the anisotropy character of the energy dispersion due to the strain; there is a symmetry breaking of the perfect circular section cuts in the conical spectrum to ellipses, with the bigger axis along the direction of strain \cite{PhysRevB.81.035411}.

\begin{figure}[!h]
\centering
\includegraphics[scale=0.38]{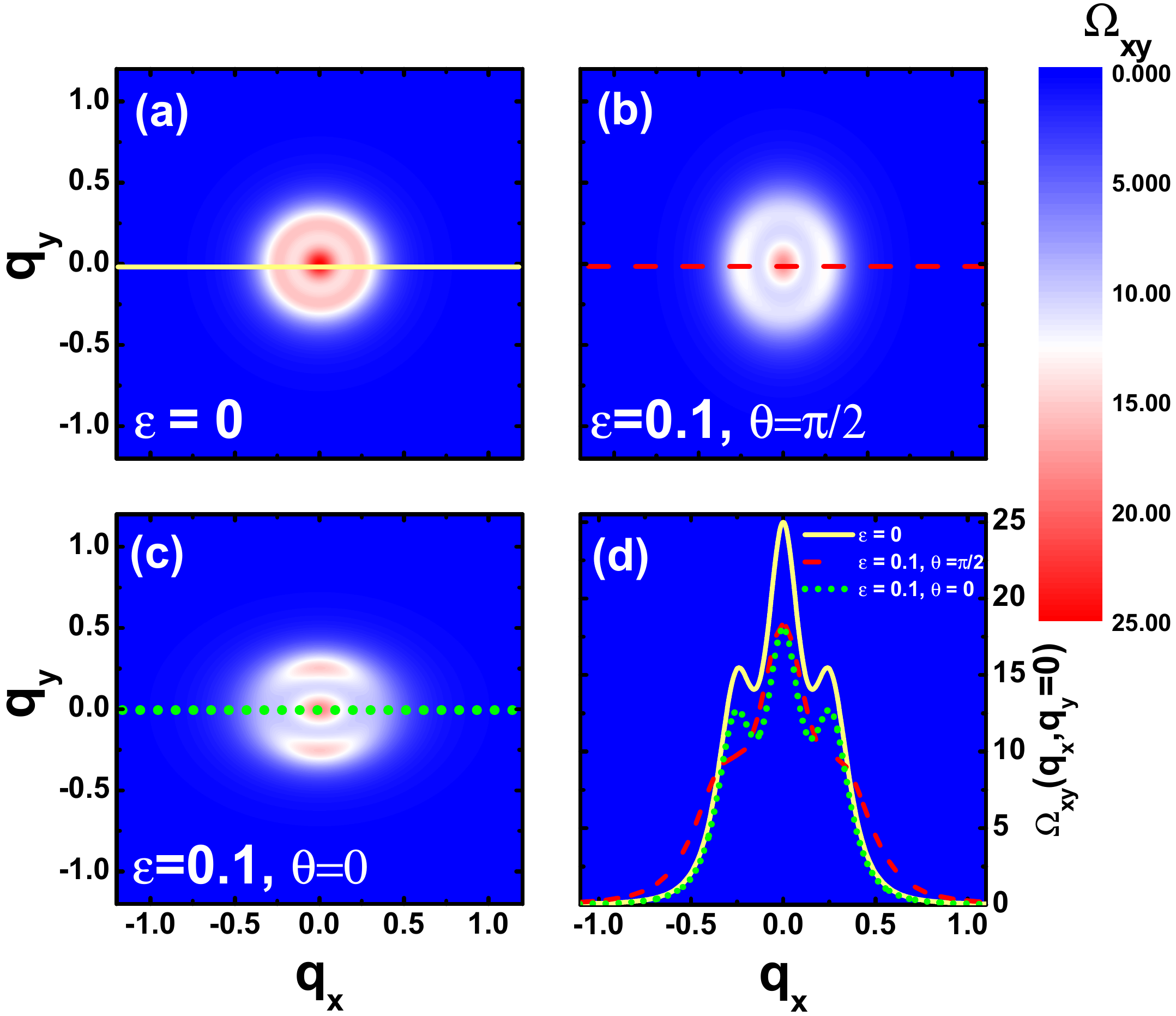}
\caption{Contour plots of the Berry curvature considering $\lambda_{R}$=0.2$t$ and $M$=0.3$t$: (a) unstrained graphene, (b) strained along $\pi/2$ with $\varepsilon$=0.1, (c) strained along $0$ with $\varepsilon$=0.1. In (d) is shown the Berry curvature for $q_{y}$=0 profile, which is associated to the lines drawn in panels (a)-(c). Notice that here we have taken the Berry curvature of the upper valence band.}
\label{grafico2}
\end{figure}

Fig. \ref{grafico2} (a)-(c) show the contour plots of the Berry curvature of unstrained (a) and strained (b), (c) graphene at the $\textbf{K}$-point for the upper valence band in the low energy approximation, with $M$=0.3$t$, $\lambda_{R}$=0.2$t$ and $\varepsilon$=0.1. We notice that the Berry curvature of the unstrained graphene shows a perfect circular symmetry in the $x$-$y$ plane. Nevertheless, it is elongated along $q_{y}$- and $q_x$- axis for the uniaxial strain being applied along $\theta$=$\pi/2$ (b) and $\theta$=0 (c), respectively. This behavior is attributed to the anisotropy of the energy bands induced by strains. In order to study quantitatively the effects of uniaxial strain on the topological phases, Fig. \ref{grafico2} (d) plots Berry curvature as a function of $q_{x}$ for $q_y$=0, as indicated by horizontal lines in Fig. \ref{grafico2} (a)-(c). We find that the uniaxial strain not only breaks down the spherical symmetry, but also reduces the maximum of Berry curvature. This behavior is strongly dependent upon the direction of strain. Besides, for a given strain size, the Berry curvature around Dirac point is less sensitive to the strain direction.

\begin{figure}[!h]
\centering
\includegraphics[scale=0.35]{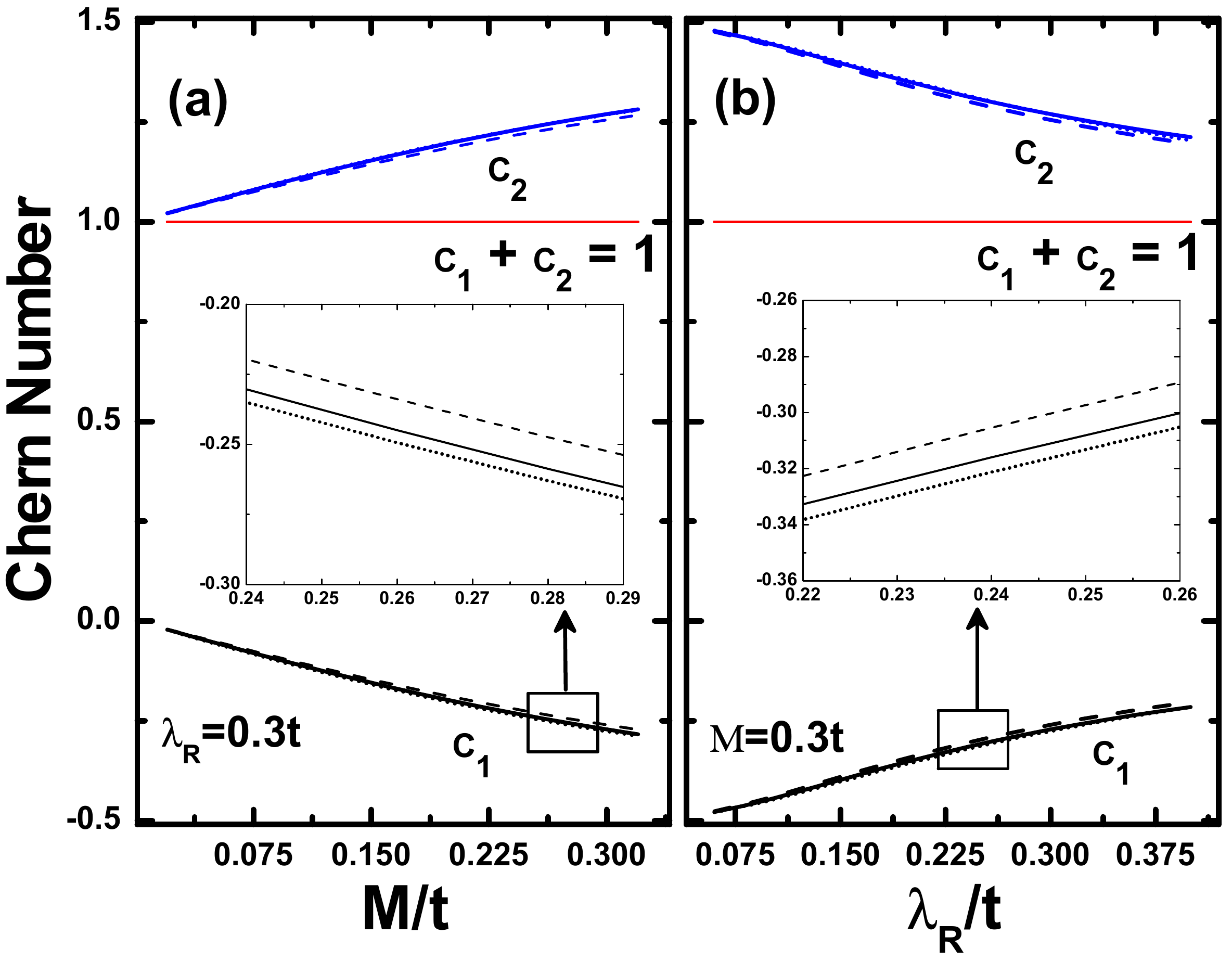}
\caption{(a) Calculated Chern number for the two valence bands as a function of $M$ for $\lambda_{R}$=0.3$t$ and different strain configurations. (b) Calculated Chern number for the two valence bands as a function of $\lambda_{R}$ for fixed $M$=0.3$t$ and different strain configurations. The inset in panels (a) and (b) shows the zoom of the squared region. Solid, dotted and dashed lines represent $\varepsilon=$0 and $\varepsilon=$0.16 for $\theta=0$ and $\theta=\pi/2$, respectively. For this calculation we have set $\lambda_{so}$=0.}
\label{grafico3}
\end{figure}

From previous calculations in unstrained graphene, it is known that graphene subjected to Rashba SOC and exchange field, both valleys contribute equally  $\mathcal{C}_{K}=\mathcal{C}_{K^{\prime}}=1$ and hence a total Chern number of $\mathcal{C}=2$ \cite{PhysRevB.82.161414}. To profoundly understand the topological property of strained graphene, we have calculated the Chern number for each valence band at $K$-point for different strength of Rashba SOC, exchange field and strain parameters. Fig. \ref{grafico3} (a), shows the dependence of Chern number on the exchange field and strain parameters for a given $\lambda_R$=0.3$t$. We notice that for the limiting case $M\rightarrow$0, the upper valence band is responsible for the Chern number $\mathcal{C}_{2}$=1, while for the lowest valence band $\mathcal{C}_{1}$=0.

Fig. \ref{grafico3} (b) plots the Chern number as a function of Rashba SOC strength for a given $M$=0.3$t$. It is noted that for Rashba SOC in the limiting case $\lambda_{R}\rightarrow$0, the upper valence band contributes to one and half quantized Chern number, while the other band contributes to minus half quantized Chern number, thus $\mathcal{C}_{K}$=1. When $\lambda_R$ increases, the absolute values of both $\mathcal{C}_{1}$ and $\mathcal{C}_{2}$ are reduced. It is also worthy to notice that although strain induces a slightly change in the slope of the Chern number of each band, the total Chern number at $\textbf{K}$ ($\textbf{K}^{\prime}$), still holds to $\mathcal{C}_{1}+\mathcal{C}_{2}=1$. Therefore, the discussion in ref. \cite{PhysRevB.85.115439} of the two limiting cases is still fulfilled in the presence of uniaxial strains, and the QAHE phase is robust under uniform lattice deformation in graphene with Rashba SOC interaction and an exchange field.

As known, ISO interaction respects the symmetries of the crystal and does not couple states of opposite spins. But it opens up a topologically non-trivial bulk band gap at zero magnetic field \cite{Kane}. This bulk band gap hosts two counter-propagating edge modes per edge in the graphene nanoribbon, with opposite spins: this topological phase is known as the QSH phase, and may be regarded as two opposite QH phases (i.e., each spin performs the QH effect, with opposite chirality) \cite{Haldane}. Therefore, the Chern number must vanish in a system with TRS. In contrast, the Rashba term explicitly violates the $z \rightarrow -z$ mirror symmetry. Moreover, it mixes different spin components and depresses the ISO induced band gap \cite{PhysRevB.81.165410}. When the exchange field is applied and only ISO is turned on, the combination of the ISO coupling and exchange field leads to the breaking of the TRS which is preserved in the QSH phase. However, due to the absence of spin-flip terms in the Hamiltonian, the helical edge-state structure persists. Thus, both the Chern number and the conductance are equal to zero. Unlikely, when Rashba SOC is considered in addition to ISO and exchange field, the system can be in a regime, which depends on $\lambda_{R}$, $\lambda_{so}$ and $M$ parameters, that may result in a phase transition from zero-conductance to finite conductance.

\begin{figure}[!h]
\centering
\includegraphics[scale=0.37]{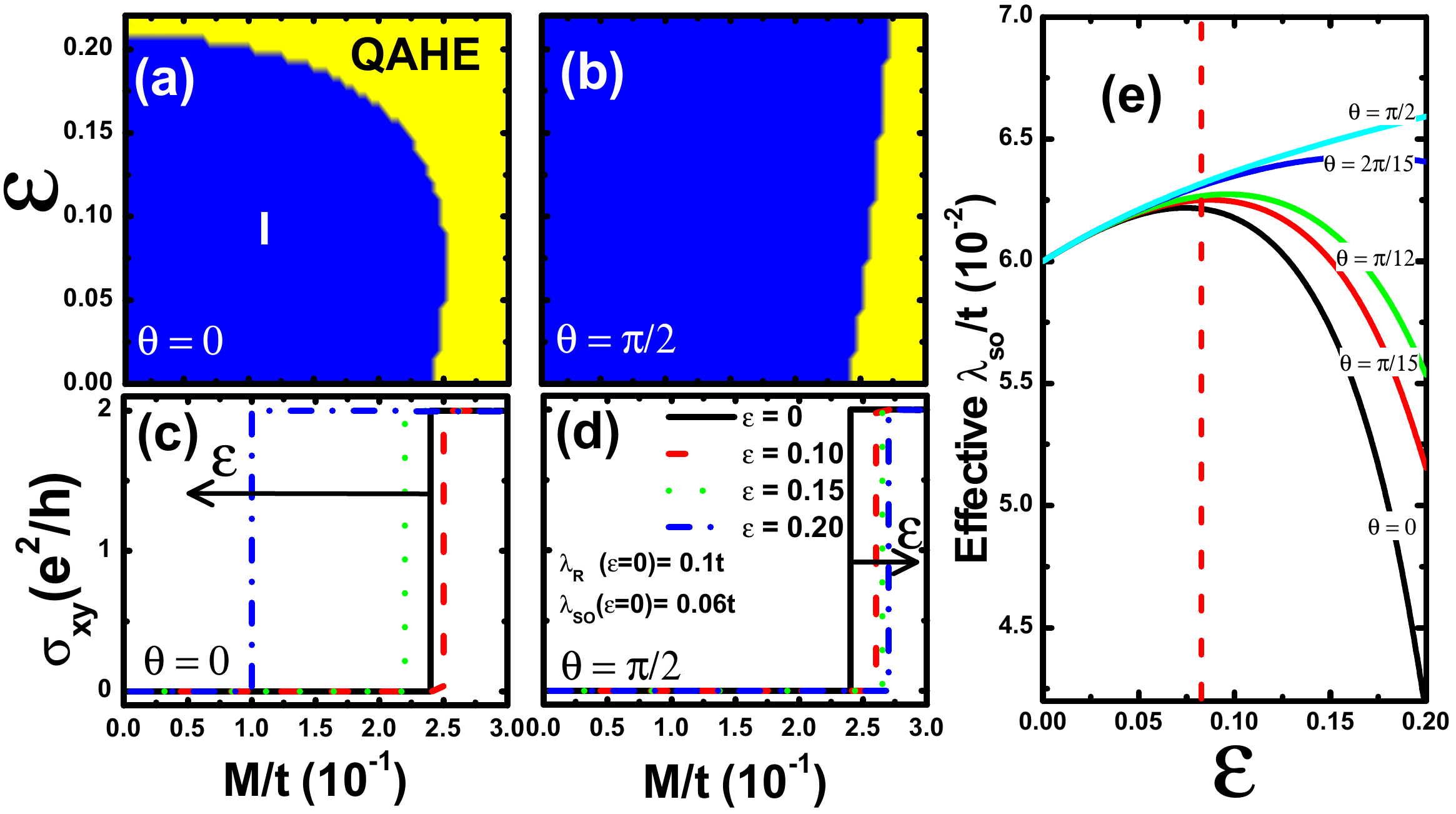}
\caption{Phase diagram of the QAHE for strained graphene along two distinct directions: (a) along $\theta=0$ and (b) along $\theta=\pi/2$. The Hall conductance as a function of the exchange interaction $M$, for uniaxial strain direction along $\theta=0$ and $\theta=\pi/2$ with four different strain strengthes is shown in (c) and (d), respectively. The arrows in panels (c)-(d) indicate the direction for which the strength is increased from $\varepsilon$=0 to $\varepsilon$=0.2. The parameters $\lambda_{R}(\varepsilon=0)=0.1t$ and $\lambda_{so}(\varepsilon=0)=0.06t$ have been used in panels (a)-(d). (e) Effective $\lambda_{so}$ as function of strain strength along different directions $\theta$. The vertical dashed line in panel (e) indicates the limiting strain modulus, for which the effective ISO parameter changes its behavior according to the direction and modulus of strain.}
\label{grafico4}
\end{figure}

Let us now calculate the Hall conductance of the strained graphene considering both Rashba SOC and ISO. Fig. \ref{grafico4} (a) and (b) show the Hall conductance for $\lambda_{R}(\varepsilon=0)=0.1t$ and $\lambda_{so}(\varepsilon=0)=0.06t$ along $\theta=0$ and $\theta=\pi/2$, respectively. One can clearly note the two distinct phases: Insulating (I) characterized by $\mathcal{C}$=0 and the QAHE phase with $\mathcal{C}$=2, where $\mathcal{C}=\mathcal{C}_{K}+\mathcal{C}_{K^{\prime}}$. The two different phases can be accessed by appropriately tuning the exchange field $M$ and the strain modulus $\varepsilon$. Fig. \ref{grafico4} (c) shows the dependence of the Hall conductance $\sigma_{xy}=\mathcal{C}e^{2}/h$ on the exchange field and the strain parameters with $\theta=\pi/2$ for $\lambda_{R}(\varepsilon=0)$=0.1$t$ and $\lambda_{SO}(\varepsilon=0)$=0.06$t$. We find that in contrary to Fig. \ref{grafico3}, in which only QAHE phase exists, a finite ISO drives a phase transition from QAHE to an insulator phase. We also notice that for $M$ being smaller than $0.24t$, the conductance $\sigma_{xy}$ of unstrained graphene is equal to zero, corresponding to an insulator phase in the graphene, also called a time-reversal-symmetry-broken quantum spin-Hall phase \cite{PhysRevLett.107.066602}. At $M_c=0.24t$, an abrupt change from $0$ to $2e^2/h$ takes place, which indicates a quantization of the Hall conductance and an occurring of a phase transition at $M=M_c$. After that, it remains $2e^2/h$, in which the unstrained graphene stays in the phase of QAHE. Furthermore, the applied strain drives Hall conductance curve forward to the right-hand side for strained graphene. Consequently, as the strain modulus increases from zero, the critical exchange field $M_c$ becomes larger, such as for $\varepsilon$=0.2, $M_c$=0.275$t$ with a relative change of $M_c$ being approximately +14.5\%. Astonishingly, in the case of $\theta=0$, as demonstrated in Fig. \ref{grafico4} (d), there is an increase in the exchange field with similar behavior for the Hall conductance. However, beyond an specific value of strain modulus, indicated by the vertical dashed line in Fig. \ref{grafico4} (e), the system presents an opposite strain-strength dependence, i.e. an increase in the strain parameter shifts the Hall conductance curve to the left-hand side. For instance, in the case of $\varepsilon$=0.2, we have obtained $M_c$=0.1$t$ with a relative change of $M_c$ being equal to -58.3\%.

The distinct behaviors observed along different strain directions for the QAHE phase transition, can be explained by the competition of the Rashba SOC and ISO in the bulk band gap-closing phenomena, for a given critical exchange field $M_{c}$ \cite{PhysRevB.84.165453,PhysRevB.85.115439}. In the case of $\theta$=$\pi/2$, an increase in the strain modulus leads to an approximately linear enhancement in the ISO parameter as can be observed in Fig. \ref{grafico4} (e), which results in an smaller bulk band gap in presence of an exchange field. On the other hand, the Rashba SOC is not very sensitive to the variation of strain strength. Therefore, the variation of Hall conductance mainly reflects the dependence of ISO on the strain strength. In contrast, for values of strain modulus larger than $\varepsilon=0.078$ in the case of $\theta$=0, there is drastic reduction in the effective ISO interaction, hence Rashba becomes dominant and the critical exchange field for the phase transition becomes smaller as one can note in Fig. \ref{grafico4} (c) with a critical $M_{c}=0.1t$ for $\varepsilon$=0.2 for the QAHE phase transition.

\section{Conclusion}
In summary, we have investigated the effects of uniaxial strains on the QAHE in graphene, by using an effective low energy approximation taking into account Rashba SOC, ISO and exchange field interaction. We show the evolution of the Berry curvature and the Chern numbers when the orientation and modulus of the uniaxial strain, as well as the exchange field interaction change. We demonstrate the tunability of Chern number and QAHE of the graphene by the strength and direction of applied uniaxial strain. In the absence of ISO, the QAHE phase is robust against the uniaxial strain. Furthermore, when ISO is considered in the system, an interesting behavior according to particular directions of strains were found: an increase of the critical exchange field $M_c$ for the QAHE phase transition for $\theta$=$\pi/2$ as the strain modulus is enhanced, in contrast to the $\theta$=0, which shows a reduction (above a limiting strain modulus of approximately $\varepsilon = 0.078$) in the critical exchange field $M_c$ for the QAHE phase. Our results suggest the possibility to efficiently manipulate the QAHE by a plausible strain engineering of the graphene structure. We envision that our study might be extended to other layered materials \cite{jz301339e,nl4013166} with potential application on novel quantum electronic devices with dissipationless charge current.

\begin{acknowledgements}
We thank fruitful discussions with F. M. D. Pellegrino and Z. Qiao. The authors also acknowledge financial support from CAPES and CNPq.
\end{acknowledgements}

\appendix
\section{Effective Hamiltonian of a Graphene Subjected to an Uniaxial Strain}
\label{alowene}
In this Appendix, we derive an effective Hamiltonian of the strained graphene in the low energy regime. After Fourier transforming Eq.\ref{H1}, the obtained $4\times4$ matrix $H(\vec{k})$ can be expanded around the strain-shifted Dirac points, $\vec{k}= \eta \textbf{K} + \vec{q}$, where $\textbf{K}=(K_{x},K_{y})$, with $\eta$=$\pm$1 related to the two valleys \cite{PhysRevB.81.035411}. The low energy Hamiltonian, $H=H_{KM}(\vec{q}) + H_M(\vec{q})$, can then be written on the basis $\{ \Psi_{A(\eta K)\uparrow}, \Psi_{A(\eta K)\downarrow}, \Psi_{B(\eta K)\uparrow}, \Psi_{B(\eta K)\downarrow}\}$ as

\begin{eqnarray}
H_{KM}(\vec{q})=\left(\begin{array}{cc}
0 & f\\
f^{*} & 0\\
\end{array}\right) + \left(\begin{array}{cc}
0 & t_{R}\\
t_{R}^{*} & 0\\
\end{array}\right)+ \left(\begin{array}{cc}
t_{so} & 0\\
0 & -t_{so}\\
\end{array}\right)
\end{eqnarray}

and

\begin{eqnarray}
H_{M}(\vec{q})=M\left(\begin{array}{cc}
S_{z} & 0\\
0 & S_{z}\\
\end{array}\right),
\end{eqnarray}

with matrix elements,

\begin{widetext}
\begin{eqnarray}
f&=&-\{t_{1}[1-i(1+\epsilon_{22})q_{y} -i\epsilon_{12}q_{x}] e^{-i\eta \epsilon_{12}K_{x}}e^{-i\eta (1+\epsilon_{22})K_{y}} \\\nonumber
     &+&t_{2}[1+i/2(\epsilon_{12}+\sqrt{3}(1+\epsilon_{11}))q_{x} + i/2(\sqrt{3}\epsilon_{21}+(1+\epsilon_{22}))q_{y}] e^{i\eta/2 (\epsilon_{12}+\sqrt{3}(1+\epsilon_{11}))K_{x}} e^{i\eta/2 (\sqrt{3}\epsilon_{21}+1+\epsilon_{22}))K_{y}} \\\nonumber
     &+&t_{3}[1+i/2(\epsilon_{12}-\sqrt{3}(1+\epsilon_{11}))q_{x} - i/2(\sqrt{3}\epsilon_{21}-(1+\epsilon_{22}))q_{y}] e^{i\eta/2 (\epsilon_{12}-\sqrt{3}(1+\epsilon_{11}))K_{x}} e^{-i\eta/2 (\sqrt{3}\epsilon_{21}-1-\epsilon_{22}))K_{y}}\}\textbf{1}_{s},\\\nonumber
\quad\\\nonumber
t_{R}&=&\lambda_{R}\{[-i(1+\epsilon_{22})e^{-2i\theta_{1}}-(\sqrt{3}\epsilon_{21}\sin{\theta_{2}}-i(1+\epsilon_{22})\cos{\theta_{2}})e^{i\theta_{1}}]S_{x} \\\nonumber
 &+&[-i\epsilon_{12}e^{-2i\theta_{1}} + (\sqrt{3}(1+\epsilon_{11})\sin{\theta_{2}}-i\epsilon_{12}\cos{\theta_{2}})e^{i\theta_{1}}]S_{y}\},\\\nonumber
\quad\\\nonumber
t_{so}&=&\eta det(\epsilon)\lambda_{so}\{2[\sin{(\sqrt{3}(1+\epsilon_{11})K_{x})}\cos{(\sqrt{3}\epsilon_{21}K_{y})} + \cos{(\sqrt{3}(1+\epsilon_{11})K_{x})}\sin{(\sqrt{3}\epsilon_{21}K_{y})}] \\\nonumber
&-& 4[\sin{(\sqrt{3}/2(1+\epsilon_{11})K_{x})}\cos{(3/2\epsilon_{12}K_{x})}\cos{(\sqrt{3}/2\epsilon_{21}K_{y})}\cos{(3/2(1+\epsilon_{22})K_{y})}\\\nonumber
&-&   \cos{(\sqrt{3}/2(1+\epsilon_{11})K_{x})}\sin{(3/2\epsilon_{12}K_{x})}\sin{(\sqrt{3}/2\epsilon_{21}K_{y})}\sin{(3/2(1+\epsilon_{22})K_{y})}]\\\nonumber
&-&   \cos{(\sqrt{3}/2(1+\epsilon_{11})K_{x})}\cos{(3/2\epsilon_{12}K_{x})}\sin{(\sqrt{3}/2\epsilon_{21}K_{y})}\cos{(3/2(1+\epsilon_{22})K_{y})}\\\nonumber
&-&   \sin{(\sqrt{3}/2(1+\epsilon_{11})K_{x})}\sin{(3/2\epsilon_{12}K_{x})}\cos{(\sqrt{3}/2\epsilon_{21}K_{y})}\sin{(3/2(1+\epsilon_{22})K_{y})}]\}S_{z}.
\end{eqnarray}
\end{widetext}

where $f$, $t_{R}$ and $t_{so}$ are the elements associated to the hopping, Rashba SOC and ISO, respectively. $\textbf{1}_{s}$ is the identity matrix, $S_{z}$ is the Pauli spin matrix in the real spin subspace, $det(\epsilon)=(1-\epsilon_{11})(1-\epsilon_{22})-\epsilon_{21}\epsilon_{12}$, $\theta_{1}=\eta/2(\epsilon_{12}K_{x} + (1+\epsilon_{22})K_{y})$ and $\theta_{2}=\sqrt{3}\eta/2((1+\epsilon_{11})K_{x} + \epsilon_{21}K_{y})$. In the unstrained limit, $\varepsilon\rightarrow0$ and $(K_{x},K_{y})=(\eta 4\pi/3\sqrt{3},0)$, it is possible to show that the low energy Hamiltonian is the same as that obtained in ref. \cite{PhysRevB.85.115439}.

\bibliography{references} 

\begin{thebibliography}{44}
\bibitem [{par()}]{parameters}%
  \BibitemOpen
  \href@noop {} {}\bibinfo {note} {Due to lattice deformation induced by the
  uniaxial strain, there is a change in the Rashba SOC and ISO interactions.
  Therefore, these interactions can effectively depend on the modulus and
  direction of strain $\lambda_{R}(\varepsilon)$ and
  $\lambda_{so}(\varepsilon)$. In the following calculations, we refer
  $\lambda_{R}(\varepsilon=0)$ and $\lambda_{so}(\varepsilon=0)$ as the values
  of SOC parameters in the unstrained limit.}\BibitemShut {Stop}%
\end{thebibliography}

\end{document}